\documentclass[epj]{svjour}
\usepackage{latexsym}
\usepackage{graphics}
\usepackage{epsfig}
\newcommand{\eqref}[1]{~(\ref{#1})}

\begin{document}

\title{$d$-wave nonadiabatic superconductivity }

\author{P. Paci$^1$, C. Grimaldi$^2$, and L. Pietronero$ ^1$}
\institute{$^1$INFM Unit\'a di Roma 1 and 
Dipartimento di Fisica, Universit\`a di Roma 1 ``La Sapienza'', 
Piazzale A. Moro 2, I-00185 Roma, Italy \\
$^2$\'Ecole Polytechnique F\'ed\'erale de Lausanne, 
DMT-IPM, CH-1015 Lausanne, Switzerland}

\date{Received: date / Revised version: date}

\abstract{The inclusion of nonadiabatic corrections to the electron-phonon
interaction leads to a strong momentum
dependence in the generalized Eliashberg equations beyond Migdal's limit.
For a $s$-wave symmetry of the order parameter, this induced momentum
dependence leads to an enhancement of $T_{\rm c}$ when small
momentum transfer is dominant. Here we study how the $d$-wave
symmetry affects the above behavior. We find that the nonadiabatic
corrections depend only weakly on the symmetry of the order parameter
provided that only small momentum scatterings are allowed for 
the electron-phonon interaction.  In this situation, We show that
also for a $d$-wave symmetry of the order parameter, the nonadiabatic corrections
enhance $T_{\rm c}$.
We also discuss the possible interplay and crossover between $s$- and $d$-wave
depending on the material's parameters.
} 
\maketitle

\section{Introduction}
In ordinary low-temperature superconductors, the smallness
of the relevant phonon frequency $\omega_0$ compared to the
Fermi energy $E_{\rm F}$ permits to formulate a theory
of superconductivity based on a closed set of formulas
known as Migdal-Eliashberg (ME) equations \cite{migdal,elia}
allowing quantitative agreements with experiments \cite{carbotte}.
The closed form of the ME equations stems from Migdal's
theorem which states that as long as $\omega_0/E_{\rm F}\ll 1$
the electron-phonon ($e$-$ph$) vertex corrections to the 
electron self-energy are at least of order
$\lambda \omega_0/E_{\rm F}$, where $\lambda$ is the $e$-$ph$
coupling, and can therefore be neglected \cite{migdal}.

A different situation is encountered in 
high-$T_{\rm c}$ superconductors such as cuprates and fullerides.
These materials have in fact Fermi energies much smaller than those
of conventional metals \cite{uemura,gunna} so that
the energy scale $\omega_0$ associated to the mediator of the 
superconducting pairing can be comparable to $E_{\rm F}$. Hence,
the quantity $\omega_0/E_{\rm F}$ is no longer
negligible and in principle vertex corrections become relevant 
preventing the ordinary ME scheme to be a correct description
of the superconducting state.

The possible breakdown of Migdal's theorem in high-$T_{\rm c}$ 
superconductors inevitably calls for a generalization beyond the ME scheme 
to include the no longer negligible vertex corrections.
A possible way to accomplish this goal is to rely on a perturbative
scheme by truncating the infinite set of vertex corrections
at a given order. In previous works, we have proposed 
a perturbative scheme in which the role of small parameter
is played roughly by $\lambda\omega_0/E_{\rm F}$ leading to
a generalized ME theory which includes the first
nonadiabatic vertex corrections \cite{GPSprl,PSG}.
For a single-electron Holstein model, such a first order
perturbative approach leads to good agreements with exact results
as long as the system is away from polaron formation, that is
for $\lambda < \lambda_c\simeq 1$ \cite{capone}. 
The region of validity of the perturbative approach, which
we could name nonadiabatic region, is characterized by 
quasi-free electrons ($\lambda < 1$)
coupled in a nonadiabatic way 
($\omega_0/E_{\rm F}$ not negligible) to the lattice.
According to this definition, our nonadiabatic region
is different from the classic polaronic picture.
Of course, larger values of $\lambda$ would render higher order vertex 
corrections important leading to the breakdown of our
truncation scheme. 

The key point of the nonadiabatic theory, is that the $e$-$ph$
effective interaction is described in terms of vertex corrections which depend
on the momentum transfer $|{\bf q}|=q$ and the Matsubara exchanged 
frequency $\omega$ in a non-trivial way.
For example, the $e$-$ph$ vertex correction appearing 
in the normal state self-energy 
becomes positive (negative) when $v_{\rm F} q<\omega$ ($v_{\rm F} q>\omega$),
where $v_{\rm F}$ is the Fermi velocity \cite{PSG}.
The generalization to the superconducting
transition reveals that this situation is also encountered for the class
of diagrams beyond Migdal's limit relevant for the Cooper channel.
Concerning the critical temperature $T_{\rm c}$,  as long as
the momentum transfer is less than $\omega_0/v_{\rm F}$, the nonadiabatic
corrections lead to a strong enhancement of $T_{\rm c}$ also for moderate
values of the $e$-$ph$ coupling $\lambda$ \cite{GPSprl,PSG,CP}.
Such a strong momentum-frequency dependence of the vertex corrections
is confirmed by numerical calculations within a tight-binding approach
\cite{pera} and general theoretical considerations on 
the physical interpretation of such nonadiabatic corrections \cite{GPSepj}.

So far, nonadiabatic superconductivity has been studied by requiring
the order parameter $\Delta$ to be independent of the momenta. 
This situation is certainly suitable
for the fullerene compounds which are $s$-wave superconductors.
However, one striking characteristic of several high-$T_{\rm c}$
superconductors is the strong momentum dependence of the order
parameter $\Delta({\bf k})$. Among the several types of measurements
aimed to resolve the pairing symmetry, the Josephson tunneling \cite{wollman}
and angle resolved photoemission experiments \cite{shen} are the
most convincing ones showing that the order parameter of several cuprates,
maybe with the exception of the electron-doped NCCO, has a predominant 
$d$-wave symmetry: 
$\Delta({\bf k})\simeq\Delta[\cos(k_x)-\cos(k_y)]$. 

The origin of the $d$-wave symmetry in high-$T_{\rm c}$ cuprates is still
debated. On one hand, the observed $d$-wave symmetry is regarded 
as an evidence against any purely electron-phonon pairing interaction
so that the mechanism 
responsible for superconductivity should be sought among pairing mediators
of electronic origin (like antiferromagnetic
fluctuations) with eventually a minor electron-phonon component. 
On the other hand, several theoretical studies have shown
that the $e$-$ph$ interaction could produce, under some
quite general circumstances, a $d$-wave symmetry of the condensate 
\cite{lichten,mierze}.
This could happen when for example charge carriers experience an
on-site repulsive interaction together with a phonon induced attraction
for large inter-electrons distances. The on-site repulsion
inhibits the isotropic $s$-wave superconducting response leading
the system to prefer order parameters of higher angular momenta.
A quite general analysis of the interplay between on-site
repulsion and neighbour and next-neighbour attraction
has shown $s$-wave to $d$-wave crossover
depending on the microscopic parameters of a model BCS Hamiltonian
\cite{fehren}.

The purpose of the present paper is to study how the $d$-wave
superconducting response resulting from a strongly momentum dependent
total interaction, is affected by the inclusion of nonadiabatic
vertex corrections. In particular, we intend to clarify whether
the complex momentum-frequency structure of the nonadiabatic  
contributions could sustain an underlying $d$-wave symmetry of
the order parameter. 

In the next section we introduce the model and the corresponding
ME equations for $s$- and $d$ wave symmetries of the gap.
In Sec. 3 we generalize the ME equations to include the
nonadiabatic terms for each symmetry channel and 
calculate the corresponding critical temperatures.
We find that the theory of nonadiabatic superconductivity can
lead to $d$-wave even for phonons in a broad parameter range which
depends on the degree of electronic correlation.

\section{The model}

In this section we introduce a simple model interaction suitable
for our investigation beyond Migdal's limit and capable of
providing for $s$- or $d$-wave symmetries of the order parameter.
Let us consider the anomalous 
self-energy at the critical temperature
\begin{equation}
\label{sigmas}
\Sigma_{\rm S}(k)=\sum_{k'}V_{\rm pair}(k-k')G(k')G(-k')\Sigma_{\rm S}(k')\ ,
\end{equation}
where $G(k')$ is the fermion dressed propagator:
\begin{equation}
\label{green}
G(k')=\frac{1}{i\omega_m-\epsilon_{{\bf k}'} -\Sigma_{\rm N}(k')}
\end{equation}
and $\Sigma_{\rm N}$ is the normal self-energy. We use the compact notation
$k\equiv({\bf k},\omega_n)$, $k'\equiv({\bf k'},\omega_m)$ and
$\sum_{k'}\equiv -T_{\rm c}\sum_m\sum_{{\bf k}'}$
where $\omega_n$, $\omega_m$ are fermionic Matsubara frequencies
and ${\bf k}$, ${\bf k}'$ are electronic momenta (from now on, all
momenta are two-dimensional vectors lying on the copper-oxygen
plane). 

To define the model interaction $V_{\rm pair}(k-k')$ we have made use 
of a number of informations gathered from previous studies. 
First, in order to obtain order
parameters with higher angular momenta than $s$-wave, it is
sufficient to consider a pair interaction made of a repulsive
part at short distances and an attractive one at higher distances
(Fig. 1).
\begin{figure}
\centerline{\psfig{figure=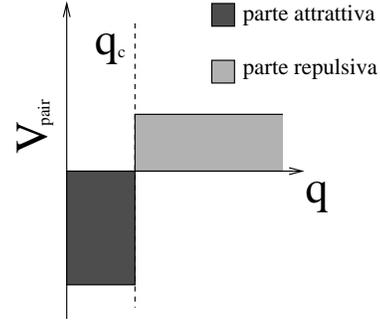,width=5cm,clip=!}}
\caption{Sketch of the total ({\em el-ph} + Coulomb) interaction
in momentum space.}
\end{figure}
In momentum space, this interaction corresponds to an attractive
coupling for small ${\bf q}$ and a repulsive one for large ${\bf q}$,
where ${\bf q}={\bf k}-{\bf k}'$ is the momentum transfer. 

Let us now try to interpret this strong momentum modulation
in terms of $e$-$ph$ and electron-electron interactions.
In strongly correlated systems, the $e$-$ph$ interaction
acquires an important momentum dependence in such a way that
for large values of the momentum transfer
${\bf q}$ the $e$-$ph$ interaction is suppressed, whereas
for small values of ${\bf q}$ it is enhanced \cite{zey}.
A physical picture to justify this momentum modulation is 
the following \cite{dany}.
In many-electrons systems a single charge carrier is surrounded 
by its own correlation hole of size $\xi$ which can be much larger than the 
lattice parameter $a$ in the strongly correlated regime. 
This implies that one electron interacts with molecular vibrations 
of wavelength of order $\xi$ or larger, leading to
an effective upper cut-off $q_{\rm c}\simeq\xi^{-1}$ in the momenta space. 
Thus we have a non zero  electron-phonon interaction when 
$|{\bf q}|<q_{\rm c}$.  The cut-off momentum
$q_{\rm c}$ can also be regarded as a measure of the correlation in the system: 
$aq_{\rm c}\ll 1$ in strongly correlated 
systems while $aq_{\rm c}\simeq 1$ in the case of free electrons.

From the above considerations, the attractive part at small ${\bf q}$
of our model pairing interaction finds a natural interpretation in the
$e$-$ph$ coupling modified by the strong electron correlations.
We introduce therefore the following simple form for the $e$-$ph$ part
of the pairing interaction:
\begin{eqnarray}
\label{coupling}
V(k-k')&=&
|g({\bf k}-{\bf k'})|^2 D(\omega_n-\omega_m) \nonumber \\
&\equiv & g^2\left[\frac{\pi k_{\rm F}}{q_{\rm c}}\right]
\theta(q_{\rm c}-|{\bf k}-{\bf k'}|) D(\omega_n-\omega_m), \nonumber \\
\end{eqnarray}
where
\begin{equation}
\label{phprop}
D(\omega_n - \omega_m)=
\frac{-\omega_0^2}{(\omega_n-\omega_m)^2+\omega_0^2}\ ,
\end{equation}
is the phonon propagator for which we have adopted a simple
Einstein spectrum with frequency $\omega_0$. In Eq.(\ref{coupling}),
$\theta$ is the Heaviside step function and
the prefactor $(\pi k_{\rm F}/q_{\rm c})$ has been introduced in order to assure that
the momentum average of  $|g({\bf k}-{\bf k'})|^2$ 
becomes $g^2$ for relatively small values of the cut-off $q_{\rm c}$ 
regardless of the particular symmetry of the order parameter.
In this way the comparison between $s$- and $d$-wave solutions,
especially in the nonadiabatic case treated in the
next section, is more transparent.

Having defined the nature of the attractive part of the total
pairing interaction $V_{\rm pair}(k-k')$, we offer now a possible interpretation
for the remaining repulsive part acting at large ${\bf q}$.
This repulsion is given by the residual {\it e}-{\it e} 
interaction and its momentum dependence can be obtained, in
analogy with the renormalization of the {\it e-ph} interaction,
by using the above picture of correlation holes.
In this picture, the residual {\it e}-{\it e}  interaction 
should ensure that charge fluctuations with wavelength less than 
$\xi$ are unfavourable. This can be modelled by requiring that in 
momentum space the residual interaction is repulsive for
$|{\bf k}-{\bf k}'|>q_{\rm c}$ and, by using again the theta-function 
for later convenience, we introduce therefore the following 
residual repulsion:
\begin{equation}
\label{u}
U({\bf k}-{\bf k'})=U\left[\frac{\pi k_{\rm F}}{q_{\rm c}}\right]
\theta(|{\bf k}-{\bf k'}|-q_{\rm c})\ .
\end{equation} 
In the above expression, $U>0$ and the factor $\pi k_{\rm F}/q_{\rm c}$ has been introduced
for the same reason as in Eq.(\ref{coupling}). Note that, in principle, 
$q_{\rm c}$ depends on $U$, however here we shall treat two quantities 
independently on each other by keeping in mind that small values of $q_{\rm c}$ 
correspond roughly to large values of $U$. 
An additional simplification of the
following calculations is achieved by expressing the off-diagonal 
self-energy (\ref{sigmas}) in terms of a suitable 
pseudopotential $U^*$ rather then $U$.
It is then opportune to formally replace equation (\ref{u}) by
\begin{equation}
\label{ustar}
U^*(k-k')=U^*\left[\frac{\pi k_{\rm F}}{q_{\rm c}}\right]\theta(|{\bf k}-{\bf k'}|-q_{\rm c})
\frac{\omega_0^2}{(\omega_n-\omega_m)^2+\omega_0^2}\ ,
\end{equation} 
where $U^*$ represents
the dynamically screened Coulomb repulsion
and
the last factor is a cut-off over the Matsubara frequencies which has been
chosen to have the same functional form of the phonon
propagator for convenience.

By summarizing the above results,  
in the off-diagonal self-energy $\Sigma_{\rm S}$, Eq.(\ref{sigmas}), 
the total pairing interaction $V_{\rm pair}(k-k')$ is given by:
\begin{equation}
\label{sigmasu}
V_{\rm pair}(k-k')=V(k-k')+U^*(k-k') ,
\end{equation}
where $V(k-k')$ and $U^*(k-k')$ are given by equation s
(\ref{coupling}) and (\ref{ustar}), respectively.
Finally, the normal state self-energy $\Sigma_{\rm N}$ entering \eqref{green} is
given by 
\begin{equation}
\label{normal1}
\Sigma_{\rm N}(\omega_n)=\sum_{k'}V(k-k')G(k') ,
\end{equation}
where the electron-electron interaction has been 
absorbed in a shift of the chemical potential.

In what follows we assume the Fermi surface to be a 
circle in the momenta space; thus the 
electronic energy $\epsilon_{\bf k}$ depends only on $|{\bf k}|$.
Moreover, we approximate equations \eqref{coupling} and \eqref{ustar}
by keeping $|{\bf k}|=|{\bf k'}|=k_{\rm F}$ so that, for example,
${\bf k}=k_{\rm F}(\cos\phi ,\sin\phi )$. In this way both $\Sigma_{\rm S}$ and
$\Sigma_{\rm N}$ depend on the momentum ${\bf k}$ only via the angle $\phi$.
At this point it is convenient to transform the momentum integrations appearing
in $\Sigma_{\rm S}$ and $\Sigma_{\rm N}$ into 
energy integrations as follows:
\begin{equation}
\label{sumint}
\sum_{{\bf k}}\rightarrow\int \frac{d\phi }{2\pi}
\int d\epsilon N(\epsilon)
\end{equation}
where $N(\epsilon)$ is the density of states for the electrons. 
We make the approximation of constant value for $N(\epsilon)=N_0$
and finite bandwidth $E$ such that the energy is defined in the 
interval $-E/2\leq\epsilon\leq E/2$. The chemical potential is $\mu=0$, 
so that we refer to the half-filled situations ($E_{\rm F}=E/2$). 

On performing the integration over the energy, the anomalous self-energy 
$\Sigma_{\rm S}$ reduces to:
\begin{eqnarray}
\label{simasu1}
&&\Sigma_{\rm S}(\phi,\omega_n)=N_0\pi T_{\rm c}\sum_{\omega_m}\int\frac{d\phi'}{2\pi}
\left[|g(\cos\theta)|^2-U^*(\cos\theta)\right] \nonumber \\
&&\times D(\omega_n-\omega_m)
\frac{\Sigma_{\rm S}(\phi',\omega_m)}{|\omega_m| Z(\phi',\omega_m)}
\frac{2}{\pi}\arctan\left[\frac{E/2}{|\omega_m| Z(\phi',\omega_m)}\right],
\nonumber \\
\end{eqnarray}
\begin{eqnarray}
\label{siman1}
Z(\phi,\omega_n)=&1&-N_0\frac{\pi 
T_{\rm c}}{\omega_n}\sum_{\omega_m}\int\frac{d\phi'}{2\pi}
|g(\cos\theta)|^2D(\omega_n-\omega_m) \nonumber \\
&\times & \frac{\omega_m}{|\omega_m| }
\frac{2}{\pi}\arctan\left[\frac{E/2}{|\omega_m| Z(\phi',\omega_m)}\right],
\end{eqnarray}
where $\Sigma_{\rm N}(\phi,\omega_n)=i\omega_n[1-Z(\phi,\omega_n)]$
and $\theta=\phi-\phi'$. The wave function renormalization
$Z(\phi,\omega_n)$ actually does not depend on $\phi$ and reduces to:
\begin{eqnarray}
\label{siman2}
Z(\omega_n)=&1&-N_0\langle g^2\rangle_0\frac{\pi T_{\rm c}}{\omega_n}\sum_{\omega_m}
D(\omega_n-\omega_m)
\frac{\omega_m}{|\omega_m| } \nonumber \\
&\times &\frac{2}{\pi}\arctan\left[\frac{E/2}{|\omega_m| Z(\omega_m)}\right],
\end{eqnarray}
where $\langle g^2\rangle_0=\int_{-\pi}^{\pi}d\theta
|g(\cos\theta)|^{2}/(2\pi)$.
Let us expand the off-diagonal self-energy \eqref{simasu1} as follows:
\begin{equation}
\label{expselfs}
\Sigma_{\rm S}(\phi, \omega_n)=
\sum_{l=-\infty}^{+\infty}\Sigma_{\rm S}^{(l)}(\omega_n)Y_l(\phi)\ ,
\end{equation}
where $Y_l(\phi)=e^{il\phi}/\sqrt{2\pi}$ are eigenfunctions
of the operator $L=-id/d\phi$. 
By requiring $\Sigma_{\rm S}(\phi, \omega_n)$ be real and invariant under
$\phi\rightarrow \phi\pm \pi$ (singlet pairing) the above 
expansion reduces to:
\begin{equation}
\label{expselfs2}
\Sigma_{\rm S}(\phi, \omega_n)=
\frac{\Sigma_{\rm S}^{(0)}(\omega_n)}{\sqrt{2\pi}}+\sqrt{\frac{2}{\pi}}
\Sigma_{\rm S}^{(2)}(\omega_n)\cos(2\phi) +\cdots ,
\end{equation}
where we have singled out the $s$-wave and $d$-wave components since
in the following we consider only these symmetries.
By multiplying both sides of \eqref{simasu1} by $Y_{l'}^*(\phi)$
and integrating over $\phi$, 
it is straightforward to show that the equations for different values 
of the index $l$ are decoupled and that $\Sigma_{\rm S}^{(l)}(\omega_n)$ reduces
to:
\begin{eqnarray}
\Sigma_{S}^{(l)}(\omega_n)=&-&N_0\left[\langle g^2\rangle_l-
\langle U^*\rangle_{l}\right]
\pi T_{\rm c}\sum_{\omega_m}
D(\omega_n-\omega_m) \nonumber \\
&\times &\frac{\Sigma_{S}^{(l)}(\omega_m)}{|\omega_m|Z(\omega_m)}
\frac{2}{\pi}
\arctan\left[\frac{E/2}{|\omega_m|
Z(\omega_m)}\right] ,
\end{eqnarray}
where
\begin{eqnarray}
\langle g^2\rangle_l&=&\frac{1}{2\pi}\int_{-\pi}^{\pi}
d\theta|g(\cos\theta)|^{2}
e^{-il\theta}
\label{g}\\
\langle U^*\rangle_l&=&\frac{1}{2\pi}\int_{-\pi}^{\pi}d\theta 
U^*(\cos\theta)e^{-il\theta}
\label{ustarmedio}
\end{eqnarray}

Finally, by introducing the coupling constants 
$\lambda_l=N_0\langle g^2\rangle_l$,
$\mu^*_l=N_0 \langle U^*\rangle_l$, and by setting 
$\Delta_l=\Sigma_{\rm S}^{(l)}/Z$, the Eliashberg
equations assume the following more familiar form:
\begin{eqnarray}
\label{siman2b}
Z(\omega_n)=&1&-\lambda_0\frac{\pi T_{\rm c}}{\omega_n}
\sum_{\omega_m}D(\omega_n-\omega_m)\frac{\omega_m}{|\omega_m|} \nonumber \\
&\times &\frac{2}{\pi}
\arctan\left[\frac{E/2}{|\omega_m|
Z(\omega_m)}\right],
\end{eqnarray}
\begin{eqnarray}
\label{simasu2}
Z(\omega_n)\Delta_l(\omega_n)=
&-&\left(\lambda_l-\mu^*_l\right)\pi T_{\rm c}\sum_{\omega_m}
D(\omega_n-\omega_m)\frac{\Delta_{l}(\omega_m)}{|\omega_m|} \nonumber \\
&\times &\frac{2}{\pi}\arctan\left[\frac{E/2}{|\omega_m|
Z(\omega_m)}\right]\ .
\end{eqnarray}

For $l=0$ and $l=2$, equations \eqref{siman2b} and \eqref{simasu2} are
Migdal-Eliashberg equations for s-wave and d-wave symmetry
channels, respectively.
The explicit expressions of the constants $\lambda_l$ and $\mu^*_l$
follow from the models we adopted for the electron-phonon and
electron-electron interactions. For $l=0$ ($s$-wave) they reduce to:
\begin{equation}
\label{avelams}
\lambda_0=\lambda\left[\frac{\pi k_{\rm F}}{q_{\rm c}}\right]
\left\langle \theta (q_{\rm c}-|{\bf k}-{\bf k'}|)
\right\rangle_{l=0}
=\lambda\frac{\arcsin Q_{\rm c}}{Q_{\rm c}},
\end{equation}
\begin{eqnarray}
\label{avemus}
\mu^*_0&=&\mu^*\left[\frac{\pi k_{\rm F}}{q_{\rm c}}\right]
\left\langle \theta (|{\bf k}-{\bf k'}|-q_{\rm c})
\right\rangle_{l=0} \nonumber \\
&=&\mu^*\left(\frac{\pi}{2Q_{\rm c}}-\frac{\arcsin Q_{\rm c}}{Q_{\rm c}}\right),
\end{eqnarray}
while for $l=2$ ($d$-wave):
\begin{eqnarray}
\label{avelamd}
\lambda_2 &=&\lambda\left[\frac{\pi k_{\rm F}}{q_{\rm c}}\right]
\left\langle \theta (q_{\rm c}-|{\bf k}-{\bf k'}|)
\right\rangle_{l=2} \nonumber \\
&=&\lambda(1-2Q_{\rm c}^2)\sqrt{1-Q_{\rm c}^2},
\end{eqnarray}
\begin{eqnarray}
\label{avemud}
\mu^*_2 &=&\mu^*\left[\frac{\pi k_{\rm F}}{q_{\rm c}}\right]
\left\langle \theta (|{\bf k}-{\bf k'}|-q_{\rm c})
\right\rangle_{l=2} \nonumber \\
&=&-\mu^*(1-2Q_{\rm c}^2)\sqrt{1-Q_{\rm c}^2},
\end{eqnarray}
where $\lambda=N_0 g^2$, $\mu^*=N_0 U^*$ and $Q_{\rm c}=q_{\rm c}/2k_{\rm F}$.
Before we generalize the above expressions to include the
nonadiabatic vertex corrections, it is useful to briefly examine
qualitatively how the the magnitude of the cut-off parameter $Q_{\rm c}$
affects the gap symmetry.
The total interaction in the gap equation (\ref{simasu2}) is weighted
by $\lambda_l-\mu^*_l$. For the $s$-wave channel it reduces to:
\begin{equation}
\label{qua1}
\lambda_0-\mu^*_0=(\lambda+\mu^*)\frac{\arcsin Q_{\rm c}}{Q_{\rm c}}-
\frac{\pi}{2Q_{\rm c}}\mu^* ,
\end{equation}
while for the $d$-wave case $l=2$ it becomes:
\begin{equation}
\label{qua2}
\lambda_2-\mu^*_2=(\lambda+\mu^*)(1-2Q_{\rm c}^2)\sqrt{1-Q_{\rm c}^2} .
\end{equation} 
When $Q_{\rm c}=1$ (that is $q_{\rm c}=2k_{\rm F}$) the repulsive interaction
(\ref{ustar}) vanishes and the $e$-$ph$ coupling (\ref{coupling}) becomes 
structureless. In this limit we expect the $s$-wave solution to
dominate over the $d$-wave one. In fact, from (\ref{qua1}) and (\ref{qua2}),
$\lambda_0-\mu^*_0=\lambda$ and $\lambda_2-\mu^*_2=0$.
By lowering $Q_{\rm c}$, the total interaction acquires a momentum dependence,
however $\lambda_2-\mu^*_2$ remains negative as long as
$1/\sqrt{2}<Q_{\rm c}<1$. In this range therefore there is not $d$-wave
solution. By further lowering of the cut-off parameter, the $d$-wave
symmetry begins to compete with the $s$-wave one and for $Q_{\rm c}\ll 1$
$\lambda_2-\mu^*_2\simeq \lambda+\mu^*$ while
$\lambda_0-\mu^*_0\simeq -\pi\mu^*/2Q_{\rm c}$ signalling that the 
$d$-wave symmetry overcomes the $s$-wave ones. 

In previous studies, we have shown that the nonadiabatic corrections
lead to an enhancement of the critical temperature for small values
of $Q_{\rm c}$ in the $s$-wave channel. However in the present model, 
small values of $Q_{\rm c}$ lead to a solution with $d$-wave symmetry
and the question we face in the following sections is whether 
also in this symmetry the nonadiabatic
corrections provide for an amplification of $T_{\rm c}$.

\section{Non adiabatic vertex corrections}

In this section we introduce the first corrections arising from the 
breakdown of Migdal's theorem
in the equations for the normal and anomalous self-energies. 
By following Ref.\cite{PSG}, the first nonadiabatic corrections
to the $e$-$ph$ interaction affect the normal state self-energy (\ref{normal1})
in the following way:
\begin{equation}
\label{normal2} 
\widetilde{\Sigma}_{\rm N}(k)=\sum_{k'}\widetilde{V}_{\rm N}(k,k')G(k') ,
\end{equation}
\begin{equation}
\label{vnsum}
\widetilde{V}_{\rm N}(k,k')=V(k-k')\left[1\! +\! \sum_{q}V(k-q)
G(q-k+k')G(q)\right] ,
\end{equation}
where the last term in the square bracket of Eq.(\ref{vnsum}) defines
the vertex function. The off-diagonal self-energy in the nonadiabatic
regime is instead modified as follows:
\begin{eqnarray}
\widetilde{\Sigma}_{\rm S}(k)&=&\sum_{k'}\left[\widetilde{V}_{\rm S}(k,k')-U^*(k-k')
\right] \nonumber \\
&\times& G(k')G(-k')\widetilde{\Sigma}_{\rm S}(k') ,
\label{sigmastilde}
\end{eqnarray}
\begin{eqnarray}
\widetilde{V}_{\rm S}(k,k')&=&V(k-k')\left[1\! +\!\sum_{q}V(k-q)G(q)
G(q-k+k')\right. \nonumber \\
&+&\left.\sum_{q}V(k-q)G(-q)G(-q+k-k')\right]\nonumber\\
&+&\sum_{q}V(k-q)V(q-k')G(q)G(q-k-k') , \nonumber \\
\label{vssum}
\end{eqnarray}
where $q\equiv({\bf q},\omega_l)$ and $U^*(k-k')$ is given by
Eq.(\ref{ustar}).
The second and the third terms within the square brackets in Eq.(\ref{vssum}) 
correspond to the first order vertex corrections, while the last term 
corresponds  to the cross scattering. 
These non adiabatic terms are shown in Fig. 2 in which we are only include 
these terms that give a finite contribution for $T=T_{\rm c}$.

\begin{figure}
\centerline{\epsfig{figure=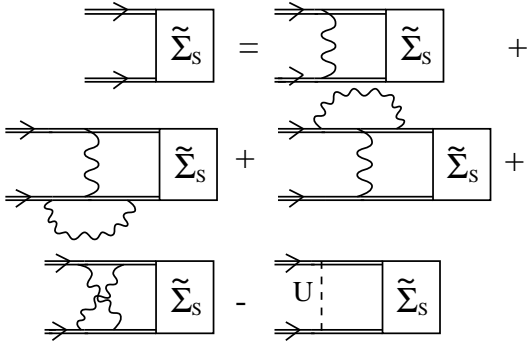,width=7cm}}
\caption{Self-consistent gap equation including the first corrections 
beyond Migdal's theorem}
\label{}
\end{figure}

In the vertex corrections there is simple one-phonon interaction, while in the 
cross term the sum refers to the product of both phonon propagators. 
This product of phonon propagators can be approximated as \cite{PSG}:
\begin{eqnarray}
&&\frac{\omega_0^2}{(\omega_n-\omega_l)^2+\omega_0^2}
\frac{\omega_0^2}{(\omega_l-\omega_m)^2+\omega_0^2} \nonumber \\
&&\simeq
\frac{\omega_0^2}{(\omega_n-\omega_m)^2+\omega_0^2}
\frac{\omega_0^2}{(\omega_n-\omega_l)^2+\omega_0^2} .
\end{eqnarray}
For the momentum dependence in the electron-phonon coupling we can approximate
\begin{equation}
|g({\bf k}-{\bf q})|^2|g({\bf q}-{\bf k}')|^2\simeq
|g({\bf k}-{\bf k}')|^2|g({\bf k}-{\bf q})|^2.
\end{equation}
This approximation is valid for relatively small values of 
the cut-off $q_{\rm c}$.
Therefore the last term in  Eq.(\ref{vssum}) reduces to:
\begin{eqnarray}
&&\sum_{q}V(k-q)V(q-k')G(q)G(q-k-k') \nonumber \\
&&\simeq V(k-k')\sum_{q}V(k-q)G(q)G(q-k-k')\ .
\end{eqnarray}
At this point it is useful to introduce a compact notation for the
vertex and cross functions: 
\begin{equation}
P_{\rm V}(k,k')\equiv
\frac{1}{\lambda}
\sum_{q}V(k-q)G(q)G(q-k+k')\ ,
\label{vertice}
\end{equation}
\begin{equation}
P_{\rm C}(k,k')\equiv
\frac{1}{\lambda}  
\sum_{q}V(k-q)G(q)G(q-k-k')\ .
\label{cross}
\end{equation}
Thus equations (\ref{vnsum}) and (\ref{vssum}) may be written in a simpler way 
as follows:
\begin{eqnarray}
\widetilde{V}_{\rm N}(k,k')&=&V(k-k')\left[1+\lambda P_{\rm V}(k,k')\right] ,
\label{vn}\\
\widetilde{V}_{\rm S}(k,k')&=&V(k-k')\left[1+2\lambda P_{\rm V}(k,k')+
\lambda P_{\rm C}(k,k')\right]. \nonumber \\
\label{vs}
\end{eqnarray}
In Eq.(\ref{sigmastilde}), the momentum dependence of $\widetilde{\Sigma}_{\rm S}(k)$
is transformed as in Eq.(\ref{expselfs}) and the interaction term
$\widetilde{V}_{\rm S}(k,k')$ is replaced by its angular weighted average:
\begin{equation}
\langle\widetilde{V}_{\rm S}(k,k')\rangle_l=
\frac{1}{2\pi}\int_{-\pi}^{\pi}d\theta\widetilde{V}_{\rm S}(\cos\theta)e^{-il\theta}\ ,
\end{equation}
which in terms of averaged vertex and cross corrections is expressed as:
\begin{eqnarray}
\label{vs1}
\langle\widetilde{V}_{\rm S}(k,k')\rangle_l&=&\langle V(k-k')\rangle_l+2\lambda
\langle V(k-k')P_{\rm V}(k,k')\rangle_l \nonumber \\
&+&\lambda\langle V(k-k')P_{\rm C}(k,k')\rangle_l  .
\end{eqnarray}
The first term in the r.h.s. contains only the $e$-$ph$
interaction and the phonon propagator and it is simply given by:
\begin{eqnarray}
\langle V(k-k')\rangle_l&=&g^2D(\omega_n-\omega_m)
\left[\frac{\pi k_{\rm F}}{q_{\rm c}}\right]
\langle \theta(q_{\rm c}-|{\bf k}-{\bf k'}|)\rangle_l \nonumber \\
&=&\frac{\lambda_l}{N_0}D(\omega_n-\omega_m),
\end{eqnarray}
where $\lambda_l$ for $l=0$ and $l=2$ is given in equations
(\ref{avelams}) and (\ref{avelamd}), respectively.
The second and third terms of (\ref{vs1}) correspond instead 
to the momentum averages of the nonadiabatic corrections
\begin{eqnarray}
\langle V(k-k')P_{\rm V}(k,k')\rangle_l 
&=&g^2D(\omega_n-\omega_m)P_{\rm V}^l(\omega_n,\omega_m,q_{\rm c}),
\nonumber \\
\langle V(k-k')P_{\rm C}(k,k')\rangle_l 
&=&g^2D(\omega_n-\omega_m)P_{\rm C}^l(\omega_n,\omega_m,q_{\rm c}),
\nonumber \\
\label{mediaver1}
\end{eqnarray}
where
\begin{eqnarray}
P_{\rm V}^l(\omega_n,\omega_m,q_{\rm c})&=&\left[\frac{\pi k_{\rm F}}{q_{\rm c}}\right]
\langle \theta(q_{\rm c}-|{\bf k}-{\bf k'}|)P_{\rm V}(k,k')\rangle_l
\nonumber \\
P_{\rm C}^l(\omega_n,\omega_m,q_{\rm c})&=&\left[\frac{\pi k_{\rm F}}{q_{\rm c}}\right]
\langle \theta(q_{\rm c}-|{\bf k}-{\bf k'}|)P_{\rm C}(k,k')\rangle_l .
\nonumber \\
\label{mediaver2}
\end{eqnarray}
Analytic expressions of the vertex and cross functions together with their
averages $P_{\rm V}^l$ and $P_{\rm C}^l$ for $l=0$ and $l=2$ are 
reported in Appendix and the results are shown in Figs. 2 and 3
as function of the adiabatic parameter $\omega_0/E_{\rm F}$ and for 
different values of dimensionless cut-off $Q_{\rm c}=q_{\rm c}/2k_{\rm F}$.
All the curves have been obtained by setting $\omega_n=0$ and
$\omega_m=\omega_0$ so that the exchanged frequency equals $\omega_0$.
The behaviors, particularly  at small $Q_{\rm c}$, of $P_{\rm V}^l$
and $P_{\rm C}^l$ are essentially independent of the particular
symmetry. In fact for both $l=0$ ($s$-wave) and $l=2$ ($d$-wave)
the nonadiabatic corrections are positive leading to an enhancement
of the total $e$-$ph$ interaction. We expect therefore that, as for the
$s$-wave case \cite{GPSprl,PSG}, also for
the $d$-wave symmetry the vertex and cross corrections tend to amplify
$T_{\rm c}$ when $Q_{\rm c}$ is sufficiently small.

\begin{figure}
\centerline{\epsfig{figure=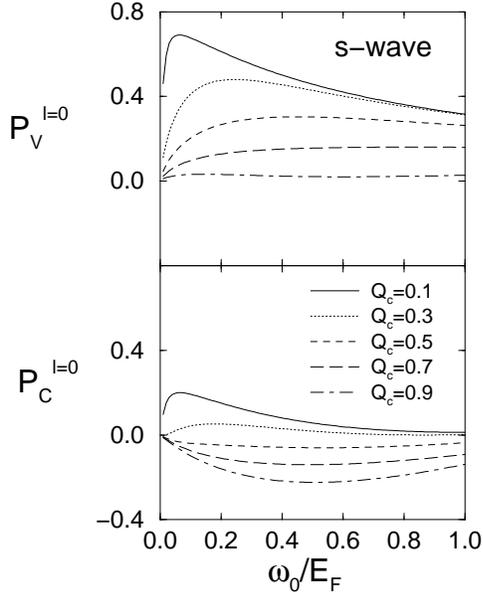,width=6cm}}
\caption{Behavior of the vertex (top panel) and cross (bottom panel) functions
in the $s$-wave channel for different values of the cut-off
parameter $Q_{\rm c}$. The case shown refers to the parameters $\omega_n=0$, 
$\omega_m=\omega_0$}
\label{s}
\end{figure}

\begin{figure}
\centerline{\epsfig{figure=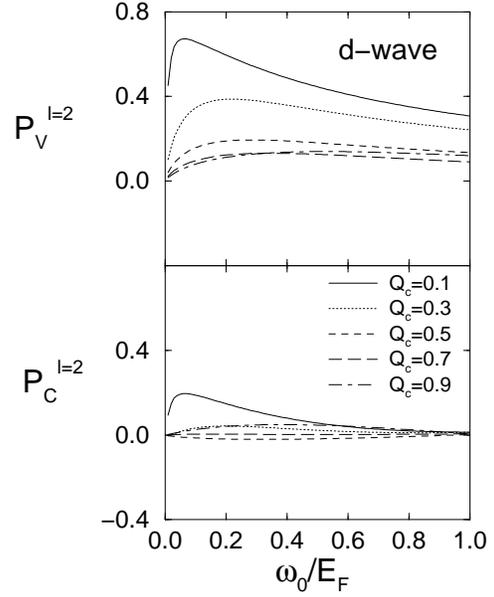,width=6cm}}
\caption{Behavior of the vertex (top panel) and cross (bottom panel) functions
in the $d$-wave channel for different values of the cut-off
parameter $Q_{\rm c}$. The case shown refers to the parameters $\omega_n=0$, 
$\omega_m=\omega_0$}
\label{d}
\end{figure}

To verify this point,
we can write down the nonadiabatic Eliashberg equations
for different symmetry channels. As in the previous section, the
normal state self-energy (\ref{normal2}) is averaged over the Fermi surface
and, according to (\ref{vn}), $\widetilde{V}_{\rm N}(k,k')$ reduces to:
\begin{eqnarray}
\widetilde{V}_{\rm N}(k,k')&\rightarrow &\langle\widetilde{V}_{\rm N}(k,k')\rangle_{l=0}\ .
\nonumber \\
&=&\frac{D(\omega_n-\omega_m)}{N_0}\left[\lambda_0+
\lambda^2 P_{\rm V}^{l=0}(\omega_n,\omega_m,q_{\rm c})\right] ,\nonumber \\
\end{eqnarray}
and $Z(\omega_n)=1-\widetilde{\Sigma}_{\rm N}^{(0)}(\omega_n)/i\omega_n$
becomes:
\begin{eqnarray}
\label{z}
&&Z(\omega_n)=1-\frac{\pi T_{\rm c}}{\omega_n}
\sum_m\left[\lambda_0
+\lambda^2 P_{\rm V}^{l=0}(\omega_n,\omega_m,q_{\rm c})\right] \nonumber \\
&&\times D(\omega_n-\omega_m)\frac{\omega_m}{|\omega_m|}
\frac{2}{\pi}\arctan\left[\frac{E/2}{|\omega_m|Z(\omega_m)}\right] .
\end{eqnarray}
Finally, the gap function for different symmetry channels is:
\begin{eqnarray}
&&Z(\omega_n)\Delta_l(\omega_n)=-\pi T_{\rm c}\sum_{\omega_m}
[\lambda_l+2\lambda ^2P_{\rm V}^l(\omega_n,\omega_m,q_{\rm c}) \nonumber \\
&&+\lambda ^2P_{\rm C}^l(\omega_n,\omega_m,q_{\rm c})-\mu^*_l]
D(\omega_n-\omega_m)
\frac{\Delta_{l}(\omega_m)}{|\omega_m|} \nonumber\\
&&\times \frac{2}{\pi}\arctan\left[\frac{E/2}{|\omega_m|
Z(\omega_m)}\right]
\label{delta}
\end{eqnarray}
where $\mu^*_l$ is given by Eqs.(\ref{avemus},\ref{avemud})
and $l=0,2$.

To establish the range of $Q_{\rm c}$ values in which the  $d$-wave 
symmetry is more stable than $s$-wave one and to quantify the
effect of nonadiabaticity, we solve numerically 
the generalized Eliashberg equations (\ref{z}) and (\ref{delta}) 
for $l=0$ and $l=2$. To find $T_{\rm c}$, we follow the maximum eigenvalue
method described for example in \cite{PSG}.
The resulting values of $T_{\rm c}$ as a function of $Q_{\rm c}$ 
are shown in Fig. 4 for the $d$- and $s$-wave symmetries.
In the inset we show the critical temperature calculated without
the vertex and cross corrections, {\it i. e.}, for the 
ME equations (\ref{siman2b}) and (\ref{simasu2}).
To display the crossover between the $d$- and the $s$-wave symmetries
more clearly, we show the results for $\lambda=1$ and $\mu^*=0.1$.
When $\mu^*\simeq\lambda$ in fact 
the $s$-wave symmetry is suppressed by the strong repulsive interaction.

\begin{figure}
\centerline{\epsfig{figure=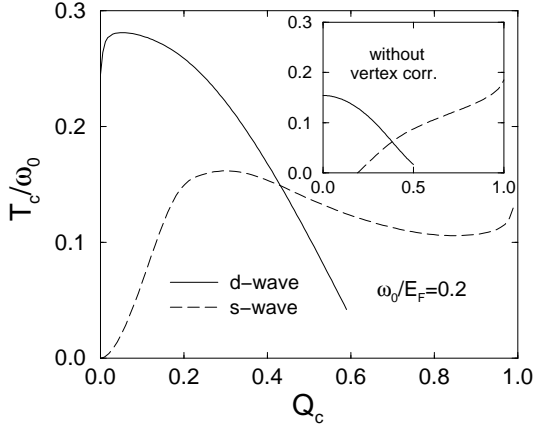,width=7cm}}
\caption{Behaviour of the critical temperature $T_{\rm c}$
as function of $Q_{\rm c}$ in the $s$- and $d$-wave 
symmetry channels. 
The case shown refers to the parameters $\lambda=1$ and 
$\mu^*=0.1$. In the inset
it is shown the case without nonadiabatic corrections.} 
\label{tc}
\end{figure}

By reducing $Q_{\rm c}$, the $s$-wave solution (dashed lines) 
decreases monotonically when the vertex and cross corrections are not
included (inset)
while for the nonadiabatic case the corresponding
$T_{\rm c}$ shows an upturn before falling to zero at $Q_{\rm c}\rightarrow 0$.
This latter feature is due to the nonadiabatic corrections which become
more positive when $Q_{\rm c}$ is small. For lower values of $Q_{\rm c}$, 
however, the pseudopotential is dominant and $T_{\rm c}$ falls
rapidly to zero.
Contrary to the isotropic case, the $d$-wave solutions (solid lines)
lead to critical temperatures which increase when $Q_{\rm c}$ is
lowered. Since, as discussed before, the vertex corrections have a similar
behavior both in $d$- and in $s$ -wave symmetries when $Q_{\rm c}$ is small,
the critical temperature in the nonadiabatic case is enhanced compared
to the solution without vertex and cross corrections.

It is finally interesting to compare the present results
with the phenomenology of the superconducting copper-oxides, 
which show $d$-wave,
and the fullerides, which instead show $s$-wave.
In our perspective there are important differencies
between the two materials. A relevant one is that the oxides
have their largest values of $T_{\rm c}$ when the Fermi surface
is strongly influenced by Van Hove singularities. Then
correlation effects can be estimated to be larger in the oxides
and, finally, fullerides seem to have rotational disorder
which would favour $s$-wave.
Therefore, in principle, it could happen that in the oxides,
going into the overdoped phase might lead to a crossover
from $d$-wave to $s$-wave depending on the parameters.

\section{Conclusions}
\label{concl}

In isotropic $s$-wave superconductors, the first nonadiabatic corrections to the $e$-$ph$
interaction such as vertex and cross functions are strongly dependent on the
momentum transfer ${\bf q}$. In particular, small values of ${\bf q}$
leads to positive nonadiabatic corrections inducing an enhancement of
the critical temperature $T_{\rm c}$ \cite{GPSprl,PSG}. 
Here, we have addressed 
the problem of the momentum dependence of the nonadiabatic corrections
for a $d$-wave symmetry of the order parameter. By introducing a model
interaction in which the $e$-$ph$ interaction is dominant at small values
of ${\rm q}$ and the residual repulsion of electronic origin is instead
important at larger momentum transfers, we have shown that also when
the solution has $d$-wave symmetry, the inclusion of nonadiabatic corrections
enhances $T_{\rm c}$ compared to the case without corrections.
Therefore in a strongly correlated system, for which the
$e$-$ph$ interaction is mainly of forward scattering, $d$-wave 
superconductivity driven by phonons can be sustained by the 
nonadiabatic corrections

\appendix
\section{Analytical calculation of vertex and cross functions}
\subsection{Vertex function}


The evaluation of the vertex function given in the Eq.(\ref{vertice}) follows 
basically the same lines and approximations made in Ref.\cite{PSG},
the main difference being that here we refer to a two-dimensional
system rather than a three-dimensional one.
Making use of the linear model for the electronic dispersion and considering 
the limit of $T_{\rm c}/\omega_0\ll 1$, we obtain
\begin{eqnarray}
&&P_{\rm V}(k,k')=
\frac{\omega_0}{2\lambda}
\sum_{{\bf q}}
\frac{|g({\bf k}-{\bf q})|^2}
{\epsilon_{\bf q}-\epsilon_{{\bf q}-{\bf k}+{\bf k}'}-i\omega_n+i\omega_m}
\nonumber\\
&&\times\left[-\frac{\theta(\epsilon_{\bf q})}
{\epsilon_{\bf q}+\omega_0-i\omega_n}-
\frac{\theta(-\epsilon_{\bf q})}
{\epsilon_{\bf q}-\omega_0-i\omega_n}\right. \nonumber \\
&&\left. +
\frac{\theta(\epsilon_{{\bf q}-{\bf k}+{\bf k}'})}
{\epsilon_{{\bf q}-{\bf k}+{\bf k}'}+\omega_0-i\omega_m}+
\frac{\theta(-\epsilon_{{\bf q}-{\bf k}+{\bf k}'})}
{\epsilon_{{\bf q}-{\bf k}+{\bf k}'}-\omega_0-i\omega_m}\right] .
\label{A1}
\end{eqnarray}
The main difficulty comes from $\epsilon_{{\bf q}-{\bf k}+{\bf k}'}$ 
which, within our model, is
\begin{eqnarray}
\epsilon_{{\bf q}-{\bf k}+{\bf k}'}&=&
v_{\rm F}[q^2+k'^2+k^2-2qk\cos{\alpha}+2qk'\cos{\beta}\nonumber \\
&-&2kk'\cos{\theta}]^{1/2}-\mu ,
\label{A2}
\end{eqnarray}
where $\alpha$, $\beta$ and $\theta$ are the angles between the directions of 
$({\bf q},{\bf k})$, $({\bf q},{\bf k'})$ and $({\bf k},{\bf k'})$ respectively.
In the limit of small $q_{\rm c}$, the presence of $\theta$-function in 
front of $P_{\rm V}$ 
[Eq.(\ref{mediaver2})] and inside of the integral leads to 
$|{\bf k}|\sim|{\bf k'}|$ and $|{\bf k}|\sim|{\bf q}|$. 
Therefore Eq.\eqref{A2} can be rewritten as follows:
\begin{equation}
\epsilon_{{\bf q}-{\bf k}+{\bf k}'}\simeq
\epsilon_{\bf q}+v_{\rm F}k_{\rm F}\left[1-\cos{\alpha}+\cos{\beta}-
\cos{\theta}\right]\ ,
\label{A3}
\end{equation}
where we have taken $|{\bf k}|\simeq k_{\rm F}$. We can relate the angle $\beta$ to 
$\alpha$ and $\theta$ by means of relation $\beta=\theta-\alpha$.
Therefore the Eq.\eqref{A3} becomes:
\begin{equation}
\epsilon_{{\bf q}-{\bf k}+{\bf k}'}\simeq
\epsilon_{\bf q}+v_{\rm F}k_{\rm F}\left[(1-\cos{\alpha})(1-\cos{\theta})+
\sin{\alpha}\sin{\theta}\right]\ .
\label{A4}
\end{equation}
If $Q=|{\bf k}-{\bf k'}|/2k_{\rm F}$ then $\cos{\theta}=1-2Q^2$ and 
$\sin{\theta}=2Q\sqrt{1-Q^2}$:
\begin{equation}
\epsilon_{{\bf q}-{\bf k}+{\bf k}'}\simeq\epsilon_{\bf q}+EQ^2(1-\cos{\alpha})+
EQ\sqrt{1-Q^2}\sin{\alpha}\ .
\label{A5}
\end{equation}
Expanding $\cos{\alpha}$ and $\sin{\alpha}$ in powers of $\alpha$ and retaining 
only the lowest order term in $\alpha$ and $Q$ we finally obtain:
\begin{equation}
\epsilon_{{\bf q}-{\bf k}+{\bf k}'}\simeq\epsilon_{\bf q}+EQ\alpha \ .
\label{A6}
\end{equation}
For small $q_{\rm c}$ we can replace
\begin{eqnarray}
\theta(q_{\rm c}-|{\bf k}-{\bf q}|)&\simeq&\theta(q_{\rm c}-k_{\rm F}\sqrt{2(1-\cos\alpha)})
\nonumber \\
&\simeq&\theta(2Q_{\rm c}-|\alpha|)\ .
\end{eqnarray}
At this point it is convenient to transform the momentum integration 
into an energy integration:
\begin{equation}
\int\frac{d^2q}{(2\pi)^2}=N_0\int_{-\pi}^{\pi}\frac{d\alpha}{2\pi}\int_
{-E/2}^{E/2}d\epsilon \,
\end{equation}
where we have used a constant DOS $N(\epsilon)=N_0$ in the range 
$-E/2\leq\epsilon\leq E/2$.
The integration over the energy $\epsilon$ and over the angle $\alpha$ 
can be performed analytically and the final expression of the vertex 
function in the limit of small $Q_{\rm c}$ is given by
\begin{eqnarray}
P_{\rm V}(k,k')&=&\omega_0B(\omega_n,\omega_m)+
\frac{\omega_0}{2Q_{\rm c}}\frac{1}{EQ} \arctan\left(
\frac{2Q_{\rm c}EQ}{|\omega_n-\omega_m|}\right) \nonumber \\
&\times &\frac{A(\omega_n,\omega_m)-
B(\omega_n,\omega_m)
(\omega_n-\omega_m)^2}{|\omega_n-\omega_m|}\ ,
\label{A10}
\end{eqnarray}
where $Q=|{\bf k}-{\bf k}'|/2k_{\rm F}$, $Q_{\rm c}=q_{\rm c}/2k_{\rm F}$,
and
\begin{eqnarray}
\label{A}
A(\omega_n,\omega_m)&=&
(\omega_n-\omega_m)
\left[\arctan\left(\frac{\omega_n}{\omega_0}\right)-
\arctan\left(\frac{\omega_m}{\omega_0}\right)\right. \nonumber \\
&+&\left.\arctan\left(\frac{\omega_m}{\omega_0+E/2}\right)
-\arctan\left(\frac{\omega_n}{\omega_0+E/2}\right)\right], \nonumber \\
\end{eqnarray}
\begin{equation}
\label{B}
B(\omega_n,\omega_m)=-(\omega_0+E/2)\frac{(\omega_0+E/2)^2+
2\omega_{m}^{2}-\omega_n\omega_m}
{\left[(\omega_0+E/2)^2
+\omega_m^{2}\right]^2}\ .
\end{equation}

\subsection{Cross function}

The function $P_{\rm C}(k,k')$, given by the Eq.(\ref{cross}), can be explicitly 
evaluated within the same scheme of calculation of the vertex function.
In the limit of $T_{\rm c}/\omega_0\ll 1$ we have:
\begin{eqnarray}
\label{B1}
&&P_{\rm C}(k,k')=
\frac{\omega_0}{2\lambda}\sum_{\bf q}
\frac{|g({\bf k}-{\bf q})|^2}
{\epsilon_{\bf q}-\epsilon_{{\bf q}-{\bf k}-{\bf k}'}-i\omega_n-i\omega_m}
\nonumber \\
&&\times\left[-\frac{\theta(\epsilon_{\bf q})}
{\epsilon_{\bf q}+\omega_0-i\omega_n}-
\frac{\theta(-\epsilon_{\bf q})}
{\epsilon_{\bf q}-\omega_0-i\omega_n}\right. \nonumber \\
&&+\left.\frac{\theta(\epsilon_{{\bf q}-{\bf k}-{\bf k}'})}
{\epsilon_{{\bf q}-{\bf k}-{\bf k}'}+\omega_0+i\omega_m}+
\frac{\theta(-\epsilon_{{\bf q}-{\bf k}-{\bf k}'})}
{\epsilon_{{\bf q}-{\bf k}-{\bf k}'}-\omega_0+i\omega_m}\right].
\end{eqnarray}
The electron energy $\epsilon_{{\bf q}-{\bf k}-{\bf k}'}$ 
can be approximated for 
$q_{\rm c}\ll 2k_{\rm F}$ as follows:
\begin{eqnarray}
\epsilon_{{\bf q}-{\bf k}-{\bf k}'}&=&v_{\rm F}[q^2+k'^2+k^2-
2qk\cos{\alpha}-2qk'\cos{\beta} \nonumber \\
&+&2kk'\cos{\theta}]^{1/2}-\mu \nonumber \\
&\simeq&\epsilon_{\bf q}+E(1-Q^2)\frac{\alpha^2}{2}-EQ\sqrt{1-Q^2}\alpha\ .
\end{eqnarray}
The integrations over the energy and the angle are elementary,
the final expression of the cross function is however
quite complicated:
\begin{eqnarray}
P_{\rm C}(k,k')&=&
\omega_0B(\omega_n,-\omega_m)-
\frac{\omega_0}{2Q_{\rm c}}
\frac{1}{E\sqrt{1-Q^2}\rho(k,k')}\nonumber\\
&\times&\left\{\cos[\eta(k,k')]
C(k,k')+\sin[\eta(k,k')]D(k,k')\right\}\nonumber\\
&\times&\frac{A(\omega_n,-\omega_m)-
B(\omega_n,-\omega_m)(\omega_n+\omega_m)^2}
{|\omega_n+\omega_m|} ,
\label{B10}
\end{eqnarray}
where the functions $A$ and $B$ are the same of Eqs.\eqref{A} and\eqref{B} 
with $\omega_m\rightarrow-\omega_m$. The function $C$, $D$, $\eta$, 
$\rho$ are given by
\begin{equation}
\rho(k,k')=\left[Q^4+
\left(2\frac{\omega_n+\omega_m}{E}\right)^2\right]^{1/4}\ ,
\end{equation}
\begin{equation}
\eta(k,k')=-\frac{1}{2}\arctan
\left(\frac{2|\omega_n+\omega_m|}{EQ^2}\right)\ ,
\end{equation}
\begin{eqnarray}
&&C(k,k')=
b(k,k')^2
\left\{  \arctan\!\left[\frac{a_+(k,k')}
{b(k,k')}\right]\right. \nonumber \\
&&+\!\left.\arctan\!\left[\frac{a_-(k,k')}
{b(k,k')}\right]\!+\!
\arctan\!\left[\frac{2Q_{\rm c}-a_+(k,k')}
{b(k,k')}\right]\right.\nonumber\\
&&+\left.\arctan\!\left[\frac{2Q_{\rm c}-a_-(k,k')}
{b(k,k')}\right]\right\},
\end{eqnarray}
\begin{eqnarray}
D(k,k')=&&
\frac{1}{2}\ln\left\{\frac{[2Q_{\rm c}-a_+(k,k')]^2+
b(k,k')^2}
{[2Q_{\rm c}-a_-(k,k')]^2+b(k,k')^2}\right. \nonumber \\
&&\times\left.\frac{a_-(k,k')^2+b(k,k')^2}
{a_+(k,k')^2+b(k,k')^2}
\right\}\ ,
\end{eqnarray}
\begin{equation}
a_\pm(k,k')=\frac{Q\pm
\rho(k,k')\cos[\eta(k,k')]}
{\sqrt{1-Q^2}}\ ,
\end{equation}
\begin{equation}
b(k,k')=
\frac{\rho(k,k')}{\sqrt{1-Q^2}}
\sin[\eta(k,k')]\ .
\end{equation}

\subsection{$s$- and $d$-wave averages}

In what follows we report the final expressions of the $s$-wave and 
$d$-wave averages of the vertex and cross functions defined 
in Eq.(\ref{mediaver2}) for $l=0$ ($s$-wave)
and $l=2$ ($d$-wave), where $P_{\rm V}$ 
and $P_{\rm C}$ are given by
Eqs.\eqref{A10} and\eqref{B10}, respectively. 
\begin{eqnarray}
&&P_{\rm V}^{l=0}(\omega_n,\omega_m;q_{\rm c})=
\frac{A(\omega_n,\omega_m)-
B(\omega_n,\omega_m)
(\omega_n-\omega_m)^2}{|\omega_n-\omega_m|}\nonumber \\
&&\times\frac{\omega_0}{2EQ_{c}^{2}}F_1(\omega_n,\omega_m,Q_{\rm c})+
B(\omega_n,\omega_m)\frac{\arcsin Q_{\rm c}}{Q_{\rm c}}
\end{eqnarray}
\begin{eqnarray}
&&P_{\rm C}^{l=0}(\omega_n,\omega_m;q_{\rm c}) \nonumber \\
&&=-\frac{A(\omega_n,-\omega_m)-
B(\omega_n,-\omega_m)
(\omega_n+\omega_m)^2}{|\omega_n+\omega_m|} \nonumber \\
&&\times\frac{\omega_0}{2EQ_{c}^{2}}F_2(\omega_n,\omega_m,Q_{\rm c})+
B(\omega_n,-\omega_m)\frac{\arcsin Q_{\rm c}}{Q_{\rm c}}
\end{eqnarray}
\begin{eqnarray}
&&P_{\ rmV}^{l=2}(\omega_n,\omega_m;q_{\rm c})=
\frac{A(\omega_n,\omega_m)-
B(\omega_n,\omega_m)
(\omega_n-\omega_m)^2}{|\omega_n-\omega_m|} \nonumber \\
&&\times\frac{\omega_0}{2EQ_{c}^{2}}
F_3(\omega_n,\omega_m,Q_{\rm c})+
B(\omega_n,\omega_m)
(1-2Q_{c}^{2})\sqrt{1-Q_{c}^{2}} \nonumber \\
\end{eqnarray}
\begin{eqnarray}
&&P_{\rm C}^{l=2}(\omega_n,\omega_m;q_{\rm c}) \nonumber \\
&&=-\frac{A(\omega_n,-\omega_m)-
B(\omega_n,-\omega_m)
(\omega_n+\omega_m)^2}{|\omega_n+\omega_m|} \nonumber \\
&&\times\frac{\omega_0}{2EQ_{c}^{2}}
F_4(\omega_n,\omega_m,Q_{\rm c})+
B(\omega_n,-\omega_m)
(1-2Q_{c}^{2})\sqrt{1-Q_{c}^{2}} \nonumber \\
\end{eqnarray}
where
\begin{equation}
F_1(\omega_n,\omega_m,Q_{\rm c})=
\int_{0}^{Q_{\rm c}}\frac{dQ}{Q\sqrt{1-Q^2}}
\arctan\left(\frac{2EQ_{\rm c}Q}{|\omega_n+\omega_m|}\right)\ ,
\end{equation}
\begin{eqnarray}
&&F_2(\omega_n,\omega_m,Q_{\rm c})=
\int_{0}^{Q_{\rm c}}\frac{dQ}{Q}
\left(\frac{1}{1-Q^2}\right)\frac{1}{\rho(k,k')}\nonumber\\
&&\times
\left\{ C(k,k')\cos[\eta(k,k')]-D(k,k')\sin[\eta(k,k')]\right\}
\end{eqnarray}
\begin{eqnarray}
F_3(\omega_n,\omega_m,Q_{\rm c})&=&
\int_{0}^{Q_{\rm c}}\frac{dQ}{Q}
\left(\frac{1+8Q^4-8Q^2}{\sqrt{1-Q^2}}\right) \nonumber \\
&\times &\arctan\left(\frac{2EQ_{\rm c}Q}{|\omega_n+\omega_m|}\right)\ ,
\end{eqnarray}
\begin{eqnarray}
&&F_4(\omega_n,\omega_m,Q_{\rm c})=
\int_{0}^{Q_{\rm c}}\frac{dQ}{Q}
\left(\frac{1+8Q^4-8Q^2}{1-Q^2}\right)
\frac{1}{\rho(k,k')}\nonumber\\
&\times&
\left\{ C(k,k')\cos[\eta(k,k')]-D(k,k')\sin[\eta(k,k')]\right\}
\end{eqnarray}
The functions $C$, $D$, $\eta$ and $\rho$ are precedently defined.

\end{document}